\documentclass[aps, prb, twocolumn, superscriptaddress, amsmath, 
tightenlines,longbibliography]{revtex4-2}

\newcommand{\RNum}[1]{\uppercase\expandafter{\romannumeral #1\relax}}
\usepackage{color}

\usepackage[utf8]{inputenc}
\usepackage[T1]{fontenc}
\usepackage[english,ngerman,danish]{babel}

\usepackage{amssymb}
\usepackage{amsmath}

\usepackage{dcolumn} 
\usepackage{graphicx}
\usepackage{mathrsfs}
\usepackage{appendix}
\usepackage{graphicx}
\usepackage{booktabs}
\usepackage{colortbl}
\usepackage{verbatim}
\usepackage{braket}

\makeatletter
\renewcommand{\fnum@figure}{%
	\textbf{Fig.~\thefigure}
}
\makeatother

\definecolor{Dred}{RGB}{190,0,0}

\usepackage{url}
\usepackage[colorlinks]{hyperref}
\hypersetup{%
	plainpages=true,
	breaklinks=true,
	hypertexnames=false, 
	pageanchor=true,
	colorlinks=true,
	linkcolor={blue},
	citecolor={red},
	urlcolor={blue},
	anchorcolor={black}
}

\usepackage{siunitx}

\usepackage{mleftright} 

\def \hide#1{}

\begin{document}
	
\title{Environmental Quantum States Trigger Emission in Nonlinear Photonics}

\author{Jia-Qi Li}
\affiliation{Institute of Theoretical Physics, School of Physics, Xi'an Jiaotong University, Xi'an 710049, People Republic of China}
	
\author{Xin Wang}\email{wangxin.phy@xjtu.edu.cn}
\affiliation{Institute of Theoretical Physics, School of Physics, Xi'an Jiaotong University, Xi'an 710049, People Republic of China}

\date{\today}


\begin{abstract}
Light-matter interactions are traditionally governed by two fundamental paradigms: spontaneous and stimulated radiation. However, in a nonlinear bath with photon-photon interactions, these paradigms break down, revealing new possibilities for photon emission. Here, we show a mechanism, termed triggered emission, in which an emitter, largely detuned from single-photon states, is triggered by the environment’s quantum state to emit a highly correlated photon pair, a doublon. By identifying two critical conditions, energy matching and wavefunction overlap, we demonstrate that the dynamics of the emitter are profoundly shaped by the environment’s quantum state. Furthermore, by engineering the initial environment photonic state and constructing a quasi-giant emitter, we realize partial unidirectional emission which evolves into a superposition state comprising a localized single-photon and a unidirectionally propagating, strongly correlated two-photon wavepacket. Our findings not only deepen the understanding of nonlinear emitter dynamics but also provide a versatile platform for quantum computing and quantum information processing.
\end{abstract}

\maketitle


\vspace{.3cm}
\noindent {\large \textbf{Introduction}}
\\
\noindent Two fundamental processes\cite{Scully1997} in quantum optics are spontaneous emission\cite{Weisskopf1930,Purcell}, where an atom emits a photon due to vacuum fluctuations, and stimulated emission\cite{EinsteinZurQD,MAIMAN1960}, in which an incoming photon promotes the emission of another identical photon. These processes have been extensively studied in linear regimes, where photon-photon interactions are significantly weaker and often negligible\cite{Haroche2006,Peyronel2012,Noh2016,Roy2017}. However, in real systems involving strong coupling\cite{Sanchez2014,Shi2018,Kuzmin2019} or multi-photon states\cite{Ke2019,Mahmoodian2020,Tomm2023}, the assumption of linearity breaks down\cite{Chang2014}. This raises an question: How do atoms behave in highly nonlinear environments, where photon-photon interactions dominate? To address this question, recent advances in nanophotonic lattices\cite{Kauranen2012,LeJeannic2022}, ultracold atoms\cite{Junemann2017,Tai2017,Antonio2020}, and superconducting circuits\cite{Yanay2020,Karamlou2024,Wang2024} have demonstrated significant enhancement of photon-photon interactions, paving the way for exploring novel radiation mechanisms. In such nonlinear regimes, photon-photon interactions fundamentally alter the emission dynamics, suggesting the emergence of new mechanisms that go beyond traditional linear optics\cite{Poshakinskiy2021,Sheremet2023,Zheng2024,Nardin2024}. 

In this work, we uncover an emission mechanism in a nonlinear photonic bath, termed triggered emission, wherein a single environmental photon triggers a far-detuned emitter to radiate. This process is distinct from spontaneous emission, which occurs without an environmental trigger. Moreover, unlike stimulated emission that produces correlated identical photon pairs, triggered emission yields two photons with distinct energies and exhibits strong spatial bunching, forming a correlated quasi-particle state ~\cite{Winkler2006,Piil2007,Mansikkamaki2022}. This mechanism highlights that in contrast to linear baths, the emission dynamics in a nonlinear bath with photon-photon interactions are governed not only by the energy spectrum but also by the quantum state of the environment~\cite{Wang2020,WangXin2024_2,Karnieli2025}.

Crucially, while environmental triggering photons can exist in diverse configurations, the generation of doublon is typically disturbed by ubiquitous single-photon transport in generic systems. To overcome this challenge, we leverage flat-band photonic lattices, a concept from condensed matter physics, where single photons are rendered immobile due to destructive interference, forming compact localized eigenstates (CLSs) \cite{Aoki1996,Leykam2018,Danieli2020,Chase2024}. Conversely, photon-photon interactions break this interference, enabling the formation of dispersive two-photon bound states (doublons)\cite{Zurita2019,Kuno2020,Flannigan2020,Pelegri2024}. These phenomena are well-established, with both theoretical predictions \cite{Vidal2000,Cartwright2018} and recent experimental verifications \cite{Zhou2023,Martinez2023,Chen2025}. 

Here, we implement triggered emission in a paradigmatic model: a two-level emitter coupled to a Bose-Hubbard lattice with synthetic gauge fields\cite{Vidal2000,Vidal2002,Cartwright2018,Danieli2021,DiLiberto2019}. This platform synergistically combines two key features: (i) Aharonov-Bohm caging enforces single-photon localization into CLSs, suppressing background single-photon transport; (ii) On-site interactions produce mobile doublons, isolating the triggered emission. Based on this platform and engineered initial single-photon states, we verify how the environmental states influence the emitter dynamics, and identify two critical conditions for triggered emission: energy matching and wavefunction overlap. Furthermore, we demonstrate its potential for generating quantum superposition states and realizing unidirectional emission, with applications in quantum information processing and integrated photonics. These results underscore the environment's decisive role in nonlinear quantum optics and open avenues for exploring novel quantum phenomena beyond the linear regime.

\begin{figure*}[t!]
	\centering \includegraphics[width=\textwidth]{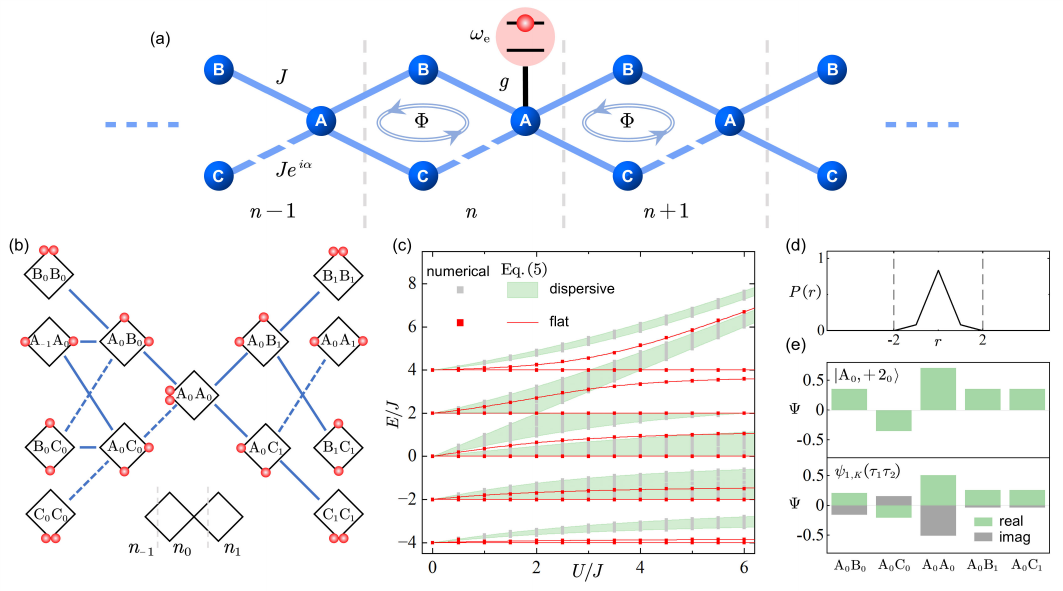} 
	\caption{\textbf{System schematic}. \textbf{a} Sketch of an excited emitter with frequency $\omega_{\text{e}}$ (red circle shading) coupled to a nonlinear rhombic chain under an external gauge field. The coupling strength is $g$ (black link). Each unit cell comprises three subsites A, B, C (blue circles), each subject to a nonlinear potential $U$. Adjacent unit cells are separated by gray dashed lines. The hopping rates for the solid (dash) links are $J$ ($Je^{i\alpha}$). The magnetic flux per rhombus is $\Phi$ and $\alpha = 2\pi \Phi/\Phi_0$. \textbf{b} Transition diagram in the two-photon space, where squares denote different Fock states and red balls represent photon positions. \textbf{c} The energy levels $E_{l}$ are obtained through numerical diagonalization of the real-space Hamiltonian Eq.~(\ref{real_space_Hamiltonian}) [gray (red) squared dots for dispersive (flat) bands], and by solving Eq.~(\ref{stationary_Schrodinger_equation}) [pale green shading (red curves) for dispersive (flat) bands]. \textbf{d} External probability distribution $P(r)=\sum_{\mu}|\psi_{K}(r,\mu)|^2$ as a function of the distance between two photons $r=|n-n^\prime|$. \textbf{e} Internal wavefunction distribution $\psi_{1,K}(\tau_1,\tau_2)$ for $E_{l=1}(K)$, with $\tau = \text{A, B, C}$. The two-photon state $\ket{\text{A}_0,+2_{0}}$ comprises one photon at $\text{A}_0$ and another photon in the eigenmode $\ket{+2_0}$. The parameters are $U=4J$ and $K=\pi/2$.}
	\label{fig1}
\end{figure*}

\vspace{.3cm}
\noindent {\large \textbf{Results and Discussion}}
\\
\noindent{\textbf{Model}}
\\
\noindent
In this article, we consider a two-level-emitter couples to a square chain lattice as depicted in Fig.~\ref{fig1}a, embedded in a uniform magnetic field with an onsite repulsively nonlinear potential $U$. The chain consists of $N$ unit cells, each containing three subsites labeled A, B, and C. Adjacent unit cells are separated by gray dashed lines.

The lattice Hamiltonian is ($\hbar=1$)
\begin{gather}
H_{\text{l}}=H_0+H_U, \label{real_space_Hamiltonian}
\\
H_0\! = \!\!-\!J\!\sum_n{\!a_{\text{A}_n}^{\dagger} \! \left(\! a^{}_{\text{B}_n} \!\! + \! e^{-i\alpha}a^{}_{\text{C}_n} \!\! + \! a^{}_{\text{B}_{n+\!1}} \!\! + \! a^{}_{\text{C}_{n+\!1}} \!\right)}\!\! +\!\! \mathrm{H}.\mathrm{c}.,
\\
H_U=\frac{U}{2}\sum_n{\sum_{\tau =\text{A,B,C}}{a_{\tau _n}^{\dagger}a_{\tau _n}^{\dagger}a_{\tau _n}a_{\tau _n}}},
\end{gather}
where $J$ is the hopping rate between the nearest-neighboring sites, and $a_{\tau_{n}}^{\dagger}$ is the photon creation operator for the subsites $\tau = \text{A,\, B,\, C}$ in the $n$th unit cell. For convenience, the gauge is chosen such that only one hopping occurs between subsites $\text{C}_{n}$ and $\text{A}_{n}$ with a phase $\alpha =2\pi \Phi /\Phi _0$. $\Phi_0 =\hbar c/e$ is the flux quantum, and $\Phi$ is the magnetic flux per square plaquette. 

The Hamiltonian for the whole system is 
\begin{gather}
H=H_{\text{l}}+\frac{\omega _{\text{e}}}{2}\sigma _z+g\left( \sigma _+a_{\tau_{n_0}}+\mathrm{H}.\mathrm{c}. \right),
\end{gather}
where $\sigma_{z,\pm}$ are the Pauli operators, $\omega_{\text{e}}$ is the transition frequency of the emitter, $\tau_{n_0}$ is the position which the emitter couples to, and $g$ is the coupling strength between the emitter and bath. For convenience, we set the frequency of the lattice sites $\omega_c = 0$ as the reference frequency throughout the main text.

\vspace{.3cm}
\noindent {\textbf{Compact dispersion of doublon bands}}
\\
\noindent
We focus on the case $\alpha=\pi$. In this scenario, the gauge field localizes a single particle in a cage, forming compact localized eigenstates (CLSs) \cite{Aoki1996} through destructive interference along two paths (upper and lower), well-known as Aharonov-Bohm caging\cite{Aharonov1959,Vidal1998,Mukherjee2018,Kremer2020,Zhang2025,Rosen2024}. The single photon spectrum is highly degenerate, and all three bands are flat with energy $\epsilon = 0,\pm 2J$. However, in the presence of nonlinearity, the particles interact with each other, forming correlated bound photon pairs, doublons. The transition diagram for two photons is illustrated in Fig.~\ref{fig1}b, with rhombuses being the Fock states in two-photon subspace and red balls indicating photon positions\cite{Kolovsky2023}. The solid (dash) lines represent single-photon hopping (with phase $\alpha$). Note that, the two-photon doublon state hops from the $\text{C}_0\text{C}_0$ to $\text{A}_0\text{A}_0$ site with a phase factor of $ 2 \times \alpha  = 2\pi$. The destructive interference is disrupted, and as a result, the two correlated photons cannot be localized by the $\pi$-flux, giving rise to the dispersive doublon bands\cite{Vidal2000,Cartwright2018,Martinez2023}.

In center-of-mass and relative coordinates, $x_c=(n+n^\prime)/2$ and $r=|n-n^\prime|$, the two photons states are expressed as $| n,\tau;n^\prime,\tau ^\prime \rangle = |x_c,r,\mu \rangle $. $\mu=(\tau,\tau^\prime)$ denotes the Fock states which two photons occupy the sublattice sites $\text{A}, \, \text{B}$, and $\text{C}$, respectively. The potential acts only at positions $(n,\tau)=(n^\prime,\tau^\prime)$, i.e., $r=0, \mu = (\tau,\tau)$. The translational invariance along $r$-direction is broken, but the $x_c$-direction is preserved. Via Fourier transformation along $x_c$-axis, the wave states can be written in the Bethe ansatz form$
\Psi \left( x_c,r,\mu \right) = e^{iKx_c}\psi _K(r,\mu )$ ,
where $K$ is the wave vector of $x_c$. Consequently, the stationary Schr\'odinger equation is derived as 
\begin{gather} 
H\Psi _{x_c,r,\mu}=E\Psi _{x_c,r,\mu},\quad 
\notag \\
\rightarrow \quad \left[ H^{\left( 2 \right)}+\delta _{r,0}H_{U}^{\left( 2 \right)} \right] \psi _K\left( r,\mu \right) =E\psi _K\left( r,\mu \right), \label{stationary_Schrodinger_equation}
\end{gather}
from which we obtain the dispersion relations $E_K$ and the wavefunction $\psi_K(r,\mu)$. The details of this equation are provided in the Supplementary Note 1. Fig.~\ref{fig1}c shows the energy spectrum as a function of $U$. The numerical results are obtained by exact diagonalization of the full two-photon subspace Hamiltonian [Eq.~(\ref{real_space_Hamiltonian})] for $N=90$, with red (gray) dots denoting the flat (dispersive) bands. The green regions (dispersive bands) and red curves (flat bands) are obtained by solving Eq.~(\ref{stationary_Schrodinger_equation}), matching well with the exact diagonalization results. Without the potential ($U=0$), the energy spectra $E=\{\pm 4, \pm 2, 0\}J$ are the linear combinations of $\epsilon=\{\pm2, 0\}J$. As $U$ increases, the original flat bands remain intact, while dispersive doublon bands emerge. For convenience, we label the bands as $E_l$, with $l=1,2,3\dots$ ordered from top to bottom.

In general dispersive systems, the nonlinear potential acts only at $r=0$ as a delta impurity in the two-photon subspace, causing an exponentially distributed wavefunction along the $r$-direction\cite{Winkler2006,Piil2007}. However, in this flat-band scenario, single photons are localized within individual cages. When two cages do not overlap and photons cannot interact, the nonlinear potential becomes ineffective. As shown in Fig.~\ref{fig1}d, we plot the modulus of the wavefunction for the first energy level $l=1$, i.e., $P(r) = \sum{_{\mu}|\psi _{1,K}(r,\mu)|^2}$. For $r>1$, $P(r)$ drops to zero. The doublon states are trapped at the position $r \leqslant 1$, two nearest-neighbor units, not exponential distribution. To simplify the description of the internal degrees of freedom, we adopt the wavefunction form $\psi_{1,K}(\tau_{n_1},\tau^{\prime}_{n_2})$, with $r=|n_1-n_2|$, and $\mu=(\tau,\tau^{\prime})$. The wavefunction of $\psi_{1,K}(\tau_{1},\tau^{\prime}_{2})$ with $\tau_1=\text{A}_0$ and $\tau_2=[\text{A}_0,\text{B}_{0/1},\text{C}_{0/1}]$ is shown in Fig.~\ref{fig1}e. The full wavefunction can be seen in Figure S1. The state exhibits maximum probability at site $\text{A}_0\text{A}_0$, with equal probabilities at sites $\text{A}_0\text{B}_{0/1}$ and $\text{A}_0\text{C}_{0/1}$. Notably, due to the chosen gauge field, the doublon wavefunction behaves oppositely for one photon located at $\text{B}_0$ and $\text{C}_0$, i.e., $\psi_{1,K}(\text{A}_0\text{B}_0)=-\psi_{1,K}(\text{A}_0\text{C}_0)$, while $\psi_{1,K}(\text{A}_0\text{B}_1)=\psi_{1,K}(\text{A}_0\text{C}_1)$, similar to the single-photon flat bands $\epsilon=\pm 2J$. 

Eventually, under the compact single photon eigenstates, the doublon states are also compact along the $r$-axis but dispersive in the $x_c$-direction , and also preserve the structure of the single-photon states. The two photons are tightly bunched within one cage and propagate through the system. 

\begin{figure}[t!]
	\centering \includegraphics[width=8.5cm]{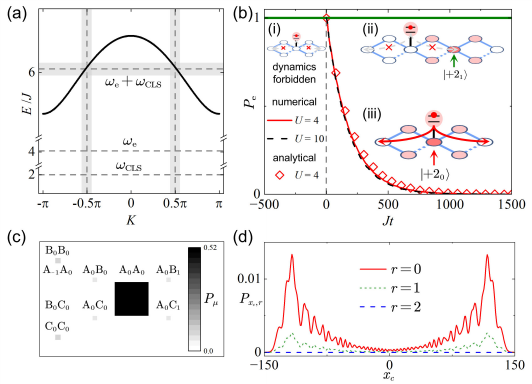}
	\caption{\textbf{Triggered emission}. \textbf{a} Energy structure of the doublon band $E_{1,K}$ (black curve). The horizontal gray dashed curves denote the emitter transition frequency $\omega_{\text{e}}$, the energy of the compact localized eigenstate (CLS) $\omega_{\mathrm{CLS}}=+2J$, and the total frequency $\omega_{\text{total}}=\omega_{\text{e}}+\omega_{\mathrm{CLS}}$. The vertical gray shaded areas indicate the resonance modes around $\pm 0.5\pi$, and the horizontal shaded areas indicate the corresponding resonance energy. \textbf{b} Time evolution of the emitter $P_{\text{e}}=|c_{\text{e}}(t)|^2$ during the triggered emission process with different nonlinear potential $U$. Before $t_0=0$, the dynamics of the emitter is forbidden [inset (i)]. At $t_0=0$, a compact localized eigenstate (CLS) is excited either at site $n_1$ [inset (ii)] or at site $n_0$ [inset (iii)], both for $U=4J$, yielding the green and red evolution curves, respectively. The black dashed curve corresponds to the case $U=10J$, with the CLS excited at $n_0$. Analytical results (red hollow diamonds) are given by Eq.~(\ref{decay_rate}). \textbf{c} Integral photon field and \textbf{d} external photon field at time $t=750J^{-1}$, plotted as functions of $\mu=(\tau,\tau^\prime)$ and $(r,x_c)$, respectively. $\tau=\text{A, B, C}$ denotes the subsite index of each unit cell $n$. $r=|n-n^\prime|$ and $x_c=(n+n^\prime)/2$ are the relative and center-of-mass coordinates, with $n, n^\prime$ being the unit cells occupied by two photons. Parameters: $g=0.02J$, $\omega_{\mathrm{CLS}}=+2J$, $U=4J$, and $\omega_{\text{e}} = 4.02J$ (for $U=4J$) or $\omega_{\text{e}} = 8.99J$ (for $U=10J$).}
	\label{fig2}
\end{figure}

\vspace{.3cm}
\noindent{\textbf{Mechanism of trigger process}}
\\
\noindent
Leveraging the flat bands and compact dispersive doublon states, we demonstrate an exotic emission process in which a CLS triggers a detuned emitter to radiate correlated photon pairs through a nonlinear resonance process. The emitter couples to the waveguide at $\tau_{n_0}$, with its frequency significantly detuned from the single-photon bands, i.e., $\omega_{\text{e}} \gg +2J$. Under this condition, the dynamics of the emitter are strongly suppressed, leaving it frozen in its excited state. The frequency setup is shown in Fig.~\ref{fig2}a. Subsequently, we excite a CLS with eigenenergy $\omega_{\mathrm{CLS}}=+2J$ at position $n_0$, i.e., the state $|+2_{n_0}\rangle$. Such an operation can be regarded as exciting a superposition state\cite{Gligoriifmmode2019,Guan2020}. Note that, we set the total energy of the emitter and CLS to lie within the first doublon band, i.e, $\omega_{\text{total}}=\omega_{\text{e}}+\omega_{\mathrm{CLS}} = E_{1,K_r}$, where $K_r$ is the resonance mode. Once the CLS is excited, the emitter can excite the doublon mode with the assistance of the CLS, reactivating previously forbidden radiation processes. Moreover, the photon emitted by the emitter combines with the CLS to form a correlated photon pair. For simplicity and without loss of generality, we set $n_0 = 0 $ and $n_{\pm1} = n_0 \pm 1 =\pm 1$. The triggered emission rate is [See detailed discussion in Supplementary Note 2]
\begin{gather}\label{decay_rate}
c_{\text{e}}\left( t \right) =e^{-\Gamma t/2},\quad \Gamma =2\frac{g^2}{v_g}|M\left( K_r, \tau_{0} ,+2_{0} \right) |^2,
\end{gather}
where $c_{\text{e}}$ is the probability amplitude for the emitter being excited state, and $v_g=\partial E_{1,K}/\partial K\mid_{K=K_r}$ is the group velocity of the first energy level $E_{1,K}$. $M$ is an effective transition rate (see Methods)
\begin{align} 
&M\left( K,\tau _n,+2_m \right) =\langle \beta _{+2_m}a_{\tau _n}\mathcal{D} _{1,K}^{\dagger}\rangle 
\notag \\
&=\!\delta _{m,n}\langle \left( 2a_{\text{A}_m}\!\!+\!a_{\text{B}_m}\!\!\!-\!a_{\text{C}_m}\!\!+\!a_{\text{B}_{m+1}}\!\!+\!a_{\text{C}_{m+1}}\! \right) \! a_{\tau _n}\! \mathcal{D} _{1,K}^{\dagger}\rangle.\!\!
\label{effective_transition_rate}
\end{align}
The delta function $\delta_{m,n}$ is non-zero only when $m = n$. This indicates that the transition is non-zero only when the excited CLS $|+2\rangle$ resides at the same position where the emitter couples. This quantity represents the overlap between the two-photon doublon mode $\mathcal{D}^{\dagger}_{1,K}|\mathrm{vac}\rangle$ and a single-photon CLS combined with single-point excitation (emitter exciting) $\beta^{\dagger}_{+2_{n_0}}a^{\dagger}_{\tau_{n_{0}}}|\mathrm{vac}\rangle$\cite{Wang2020,WangXin2024_2}. $\beta^{\dagger}_{(0,\pm2)_{n}}$ is the creation operator for the CLS in eigenmode $\epsilon = 0,\pm2$ located at site $n$ .

In traditional linear single-emitter radiation, the decay rate is proportional to $\langle a_n \Psi(x)\rangle$, the overlap between the single-point state excited by the emitter $\sigma _- a_{n}^{\dagger}|e\rangle$, and the wavefunction of the resonance mode, $|\Psi(x)\rangle$. In this process, the emitter can independently excite the single-photon state. Within two-photon subspace, the emitter only contributes a single-point excitation $\sigma _- a_{\tau_{n}}^{\dagger}|e\rangle$. To excite the doublon state $\mathcal{D}_{1,K}^{\dagger}|\mathrm{vac} \rangle$, it must combine with another single-photon state $\beta^{\dagger}_{+2_m}|\mathrm{vac} \rangle$. Consequently, the dynamics of the emitter depend not only on the photon state wavefunction, but also on the photon state of the system, which is absent in conventional spontaneous emission. Moreover, the radiation field  consists of a quasi-particle doublon formed by two photons with distinct  energies, representing a process fundamentally different from stimulated emission. The doublon mode can be also excited using two emitters coupled to the waveguide~\cite{Wang2020,WangXin2024_2}. One emitter first supplies an intermediate single-photon state, which then enables the other emitter to excite the doublon, yielding a radiation rate $\Gamma\propto g^4$. In contrast, our proposal replaces one emitter with an environmental single-photon state, which directly cooperates with the emitter to jointly excite the doublon mode, resulting in an enhanced decay rate scaling as $\Gamma \propto g^2$.

In Fig.~\ref{fig2}b, we plot the population of the emitter through numerical simulations of the entire system with waveguide $N=300$, along with analytical results given by Eq.~(\ref{decay_rate}). In the first part, the dynamics are suppressed due to no available modes exist for the emitter to excite, leaving it frozen in its excited state. However, owing to the nonlinear potential, doublon states emerge, providing new decay channels. Once the CLS $|+2_{0}\rangle$ is excited at $t=0$, the emitter cooperates with the CLS to excite the doublon mode, with exponential emission. To illustrate the two-photon field, we present a bubble chart in Fig.~\ref{fig2}c to display the integral degree $\mu$ and photonic filed in Fig.~\ref{fig2}d to visualize the external degrees $x_c,r$. The size and color of the bubbles represent the field distribution $P_{\mu}=\sum{_{x_c}|\psi (x_c,r,\mu)|^2}$, mapping to the Fock state shown in Fig.~\ref{fig1}b. Additionally, the 3D wall plot depicts the field distribution $P_{x_c,r}=\sum{_{\mu}|\psi (x_c,r,\mu)|^2}$. Most strikingly, due to the CLS and the compact doublon state, the two photons are tightly bunched within the cage regime $r<1$, and exhibit maximum probability at the $\text{A}_0\text{A}_0$ position, showcasing super-correlated characteristic. Initially, the two photons are uncorrelated (one in the emitter and one in the CLS), but the triggered emission process correlates them.

Moreover, due to the compact property, increasing the nonlinearity strength induces only minor changes in the doublon wavefunctions. According to Eq.~(\ref{decay_rate}), the decay rate is proportional to the wavefunction. Consequently, it also undergoes minimal changes. We plot the evolution of emitter with $U=10J$ in Fig.~\ref{fig2}b, maintaining $\omega_{\text{total}} = \omega_{\text{e}} + \omega_{\mathrm{CLS}} = E_{1,K_r}$. As shown in Fig.~\ref{fig1}b and Fig.~\ref{fig2}a, which detail the energy structure of the doublon bands and the trigger setup, the emitter frequency exhibits significant detuning from single-photon scattering under the condition $U=10J$, with $\omega_{\text{e}} = 8.99J\gg+2J$. Despite this, the emitter remains capable of photon emission, with only slight variations in the decay rate compared to the case of $U=4J$.

For a nonlinear cavity, the presence of one photon induces a frequency shift, which renders the entry of additional identical photons energetically unfavorable, thereby resulting in photon blockade \cite{Chen2022}. In our proposal, when a photon is localized within a cage, the nonlinear potential also causes energy level shifts. As a result, an excited emitter within this cage cannot excite single-photon scattering states or emit photon spontaneously on its own, which is fundamentally distinct from conventional spontaneous emission. Instead, the emitter can only hybridize with another single-photon state to excite two-photon doublon states, induced by $U$. Moreover, the emitted photons differ in energy from the triggering photon and together form a correlated pair that behaves as a quasi-particle exhibiting strong photon bunching, which is also fundamentally different from stimulated emission. In the single-photon regime, the excited emitter radiates into the vacuum field. While in the nonlinear bath, the emitter can mediate the transition between the single-photon state and two-photon state.

\begin{figure}[t!]
	\centering \includegraphics[width=8.5cm]{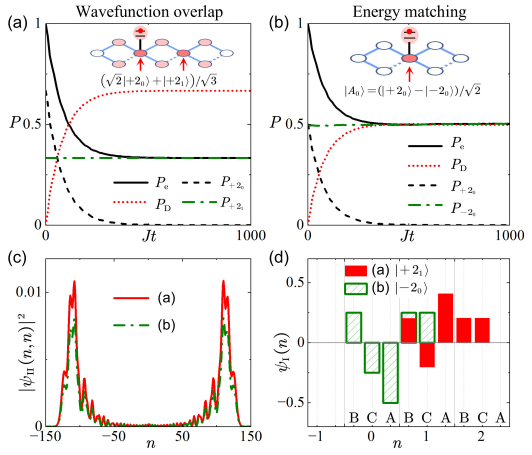}
	\caption{\textbf{Two conditions for triggered emission}. \textbf{a} Wavefunction overlap condition and \textbf{b} energy matching condition. The populations of the emitter ($P_{\text{e}}$, black solid curves), the doublon state ($P_{\text{D}} = \sum_{m,n}{|c_{m,n}|^2}$, red dotted curves), and three compact localized eigenstates (CLSs) ($P_{+2_0}$, black dashed curve; $P_{+2_1}$ and $P_{-2_0}$, green dash-dotted curves) are shown. The excited CLSs $(\sqrt{2}\ket{+2_0}+\ket{+2_1})/\sqrt{3}$ and $\ket{A_0}=(\ket{+2_0}-\ket{-2_0})/\sqrt{2}$ are depicted in the insets of panels (\textbf{a}) and (\textbf{b}), respectively. \textbf{c} Two-photon field distribution versus $n$ at time $t=700J^{-1}$ and $r=0$, given by $|\psi(n,n)|^2 =\sum_{\tau} |\psi(\tau_n,\tau_n)|^2$. \textbf{d} Single-photon localized state distribution $\psi(\tau_n)$, where $\tau=\text{A, B, C}$ labels the subsite index of each unit cell $n$. The red and green curves/bars represent the field distributions corresponding to conditions (\textbf{a}) and (\textbf{b}). The coupling strength is $g=0.03J$ and other parameters are the same as in Fig.~\ref{fig2}.}
	\label{fig3}
\end{figure}

Based on the trigger mechanism, we present two paradigmatic cases to clarify the conditions for emission: wavefunction overlap and energy matching. 

\vspace{.3cm}
\noindent {\small\textbf{Wavefunction overlap}}
\\
\noindent
The occurrence of emission requires the effective transition rate, given by Eq.~(\ref{effective_transition_rate}), to be non-zero, i.e., $M\ne 0$, which implies the two two-photon wavefunctions exhibit spatial overlap in real space. We excite the CLS in the nearest-neighbor cell of the emitter's coupling site, i.e., $n_1=n_0+1=1$, and the effective transition is 
\begin{align}\label{n_0_n_1}
&M\left( K,\text{A}_0,+2_{1} \right) 
\notag \\
&=\langle \left( 2a_{A_{1}}+a_{\text{B}_1}-a_{C_{1}}+a_{\text{B}_{2}}+a_{C_{2}} \right) a_{\text{A}_0}\mathcal{D} _{1,K}^{\dagger}\rangle .
\end{align}
As shown in Fig.\ref{fig1}d, the compact dispersion state exhibits no wavefunction distribution at position $\text{A}_0A_1$, $\text{A}_0\text{B}_2$ and $\text{A}_0C_2$ ($r>1$). Their contributions to the dynamics can be neglected. The transition is approximate to 
\begin{gather}
M\left( K,\text{A}_0,+2_{1} \right) \! \simeq \!\langle \left( a_{\text{B}_1}-a_{C_{1}} \right)\! a_{\text{A}_0}\mathcal{D} _{1,K}^{\dagger}\rangle\! = \! 0.
\end{gather}
Fig~\ref{fig1}e shows that the doublon wavefunction distribution is identical for $\psi_{1,K}(\text{A}_0\text{B}_1)=\psi_{1,K}(\text{A}_0\text{C}_1)$, but the excited CLS has opposite phases at $\text{B}_1$ and $\text{C}_{1}$, resulting in a zero transition amplitude. Therefore, the overlap between two two-photon states vanishes. Conversely, when the CLS is located to the left of the emitter, i.e., $n_{-1} = n_0 - 1 = -1$, the transition rate remains zero. Unlike in the previous section, this scenario occurs because the phases of the CLS at $\text{B}_{0}$ and $\text{C}_{0}$ are identical, while the phases of the doublon states are reversed $\psi_{1,K}(\text{A}_0\text{B}_0)=-\psi_{1,K}(\text{A}_0\text{C}_0)$, resulting in a zero transition rate. As shown in Fig.~\ref{fig1}b (green curve), we excite a CLS $|+2_{1}\rangle$ nearest to the emitter, and the triggering process no longer occurs, leaving the emitter still frozen. 

Furthermore, we excite a superposition state, involving two CLSs at different positions
$$|\psi \left( t=0 \right) \rangle =\gamma _0|e,+2_{0}\rangle +\gamma _1|e,+2_{1}\rangle,$$
where $\gamma_{0/1}$ are the amplitude coefficients of two states, satisfying the normalization condition $\gamma _{0}^{2}+\gamma _{1}^{2}=1$. Fig~\ref{fig3}a shows the population of the emitter ($P_{\text{e}}$), doublon state ($P_{\text{D}}$), and the two CLSs ($P_{+2_{0/1}}$) for $\left( \gamma _0=\sqrt{2},\gamma _1=1 \right) /\sqrt{3}$. The emitter selectively combines with the CLS $|+2_{0}\rangle$ at $n_0$ to excite the doublon state, while the other CLS $|+2_{1}\rangle$ at $n_1$ remains localized, forming the final state
\begin{gather}
|\psi \left( t_{\text{f}} \right) \rangle =\gamma _0|g,\psi _{\exp}\rangle +\gamma _1|e,+2_{1}\rangle ,
\end{gather}
which is a superposition consisting of an exponentially propagating two-photon correlated wave packet $|\psi_{\exp}\rangle $, and a single-photon localized state. Due to the compact property of the system, the excited single photon must reside within the same cage (i.e., the CLS region) where the emitter is coupled, thereby triggering the emitter's radiation. 

\vspace{.3cm}
\noindent {\small\textbf{Energy matching}}
\\
\noindent
This process also satisfies the law of energy conservation, as the sum of the CLS energy and the emitter frequency resonates with the doublon bands. We excite a single point $\text{A}_0$, to which the emitter couples. The initial state in real space and the eigenmode space is 
$$ |\psi \left( t=0 \right) \rangle = |e\rangle \otimes |\text{A}_0\rangle =\frac{1}{\sqrt{2}}\left( |e,+2_{0}\rangle -|e,-2_{0}\rangle \right).$$ 
Note that, it naturally forms a superposition state between two CLSs with different energy. We still set $\omega_{\text{e}} + 2J = E_{1,K_r}$. As discussed above, the emitter combines $|+2_{0}\rangle $ state to excite the doublon mode, radiating an exponential wave packet $|\psi_{\exp}\rangle $. However, the energy $\omega_{\text{e}} - 2J$ does not tune with any doublon mode, and the state $|-2_{0}\rangle $ remains localized. Fig~\ref{fig3}b shows the population evolution of the emitter and two CLSs $P_{\pm 2_{0}}$ under the condition $\omega_{\text{e}}+2J=E_{1,K_r}$. Following the triggered emission process, the system forms a superposition state consisting of a localized state and an exponentially propagating wave packet,
\begin{gather}
|\psi \left( t_{\text{f}} \right) \rangle = \frac{1}{\sqrt{2}}\left( |g,\psi _{\exp}\rangle -|e,-2_{0}\rangle \right).
\end{gather}
We can also set $\omega_{\text{e}} - 2J = E_{1,K_r}$, such that the emitter combines with $|-2_{0}\rangle $ state for radiation, while $|+2_{0}\rangle $ remains localized. The final state is $|\psi \left( t_{\text{f}} \right) \rangle = \left( |e,+2_{0}\rangle -|g,\psi _{\exp}\rangle \right) / \sqrt{2}$.

Fig~\ref{fig3} c,d show the single- and two-photon field distributions for both scenarios. The CLS that does not satisfy the overlap and energy conditions remains localized at its excitation position. In contrast, the CLS satisfying these conditions combines with the photon emitted by the emitter to form  a correlated two-photon pair, which propagates along the bath. 

More generally, based on this setup, if an arbitrary single-photon state is excited, the system naturally evolves into a superposition state consisting of a localized state and a propagating doublon wave packet. An arbitrary single photon state can be expressed in both real and eigenmode space as  
\begin{align}
|\psi \left( t_0 \right) \rangle =&\sum_n{\sum_{\tau =\text{A,B,C}}{f\left( n,\tau \right) |n,\tau \rangle \otimes |e\rangle}}
\notag \\
=&\sum_n{\sum_{\beta =0,\pm 2J}{g\left( n,\beta \right) |n,\beta \rangle \otimes |e\rangle}},
\end{align}
where $f(n,\tau)$ and $g(n,\beta)$ are the probability amplitudes of the excited state localized at position $(n,\tau)$ with eigenmode $\beta$, respectively. The transition matrix between $f$ and $g$ corresponds to the wavefunction of the single-photon [see
Supplementary Note 3]. The final superposition state is
\begin{align}
&|\psi \left( t_{\text{f}} \right) \rangle =g\left( n_{\text{e}},\beta _{\text{e}} \right) |g,\psi _{\text{D}}\left( l,K_r \right) \rangle 
\notag \\
&+\sum_n{\sum_{\beta =0,\pm 2J}{g\left( n,\beta \right) |n,\beta \rangle \otimes |e\rangle \mid _{_{n\ne n_0\,\,\&\beta \ne \beta _0}}}},
\end{align}
where $n_{\text{e}}$ and $\beta_{\text{e}}$ represent the position and energy of the CLS which satisfies the overlap and energy condition. $|\psi_{\text{D}}(l,K_r)\rangle $ denotes the propagating two-photon doublon packet with energy $E_{l,K_r}$, where $l$ indicates the energy level and $K_r$ is the tuning mode. The latter terms are the still localized single-photon states. 

This proposal can generate a superposition state between a localized state and a mobile two-photon packet. As shown in Fig.~\ref{fig1}b, the rich structure of the doublon bands provides extensive degrees of freedom for the final superposition state. The ratio of the mobile wave packet to the localized state, the shape and the energy of the wave packet, and the trigger single-photon CLS can all be modulated, offering versatile control for encoding and transmitting diverse quantum information.

\begin{figure}[t!]
	\centering \includegraphics[width=8.5cm]{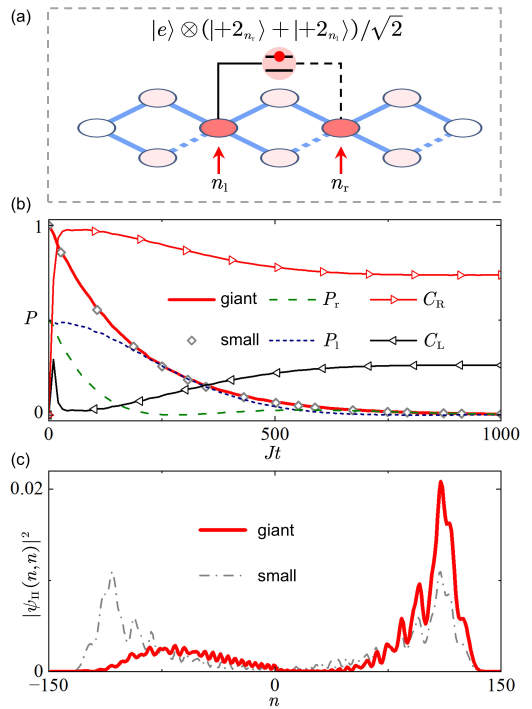}
	\caption{\textbf{Quasi-emitter applications}. \textbf{a} The setup for a quasi-giant emitter. The coupling strength between the giant emitter and the position $n_{\text{r}}$ ($n_{\text{l}}$) is $ge^{i\phi}$ ($g$), denoted by dashed link (solid link), respectively.  \textbf{b} Populations of the small emitter (gray hollow diamonds), the quasi-giant emitter (red solid curve), two compact localized eigenstates (CLSs) $\ket{+2_{n_{\text{r}}}}$ ($P_{\text{r}}$, black dotted curve) and $\ket{+2_{n_{\text{l}}}}$ ($P_{\text{l}}$, green dashed curve), and two chiral factors ($C_{\mathrm{R}}$, red curve with right-pointing triangle; $C_{\mathrm{L}}$, black curve with left-pointing triangle). \textbf{c} Photon field of the small emitter (gray dash-dotted curve) and the quasi-giant emitter (red solid curve) at time $t=700J^{-1}$. $\phi=-\pi/2$ and the parameters are the same as in Fig.~\ref{fig2}.
	}
	\label{fig4}
\end{figure}

In light of the above criteria, the supercorrelated radiation from emitter pairs~\cite{Wang2020,WangXin2024_2}, can be re-examined. Such radiation arises only when two conditions are simultaneously satisfied: the total frequency of the pair lies within the doublon band, and their spatial separation is contained within the spatial extent of the doublon wavefunction. Once these conditions are met, this process proceeds autonomously, independent of the initial environment photonic state.

In contrast, the trigger emission mechanism demonstrates that a single photon present in the environment can trigger the radiation from a frozen emitter. Although the single photon and the emitted photon differ in energy and origin, they can still hybridize into a quasi-particle, doublon. For an arbitrary single-photon state, only the component which satisfies both conditions, can trigger partial emission from the emitter. Crucially, this selective interaction naturally leads to the entangled superposition states between the emitter and the photonic field. The presence of an environmental photon introduces an additional degree of freedom for quantum control.

\vspace{.3cm}
\noindent {\textbf{Quasi emitter}}
\\
\noindent
In condensed matter physics, excitons are formed through the binding of a hole and an electron\cite{Schfer2002}. Analogously, the CLSs and emitters can combine to act as a quasi emitter, 
emitting correlated photon pairs. In our proposal, the two photons reside in distinct subspace: the environment bath and emitter. The radiation photon field is composed of super-correlated photons, providing a platform for encoding more quantum information. These characteristics offer additional degrees of freedom to explore intriguing phenomena, and extend applications from linear to nonlinear regime, such as unidirectional emission\cite{Lodahl2017}. 

\vspace{.3cm}
\noindent {\small \textbf{Triggered unidirectional radiation}}
\\
\noindent
In the linear regime, unidirectional radiation can be realized through two setups: (1) a giant emitter coupled to the waveguide at multiple points $A_{i}$\cite{Anton2014,Kannan2020,Gong2024}; (2) two entangled emitters coupled to the waveguide at $A_{i}$\cite{Guimond2020}. Both configurations have been experimentally demonstrated for entanglement states and related applications, providing a foundation for extending these concepts into the nonlinear regime~\cite{Joshi2023,RN519}. In this paper, we take the giant-atom configuration as an example to demonstrate chiral emission. A similar setup can also be employed for entangled emitter pairs to achieve chirality. Owing to the characteristics of quasi emitter, it is also necessary to excite the CLSs $|+2_{i}\rangle$ at the corresponding locations to form a quasi-giant emitter. The interaction Hamiltonian and the initial state are 
\begin{gather}
H_{\mathrm{int}}=\sum_i{ge^{i\phi _i}\sigma _-a_{{\text{A}}_{i}}^{\dagger}}+\mathrm{H}.\mathrm{c}.,\quad 
\\
|\psi(t=0)\rangle=|e\rangle \otimes  \sum_i{\frac{|+2_i\rangle}{\sqrt{N}}} . \label{psi_full}
\end{gather}
As shown in Fig.~\ref{fig4}a, the emitter couples to the waveguide at two points $n_{\text{r}},n_{\text{l}}$, where $n_{\text{l}}-n_{\text{r}}=d > 0$. The frequency is still set as $\omega_{\text{total}} = \omega_{\text{e}} + \omega_{\mathrm{CLS}}= E_{1,K_r}$. Since Eq.~(\ref{effective_transition_rate}) is nonzero only for $m=n$, the CLS at $n_{\text{r}}$ ($n_{\text{l}}$) cannot trigger the leg located at $n_{\text{l}}$ ($n_{\text{r}}$), i.e., $$M\left( K,\text{A}_{n_{\text{r}}},+2_{n_{\text{l}}} \right) =M\left( K,\text{A}_{n_{\text{l}}},+2_{n_{\text{r}}} \right) =0.$$
The effective transition rate is thus simplified as 
\begin{align}
M&\left( K,\text{A}_{n_{\text{r}}},+2_{n_{\text{r}}} \right) +e^{i\phi}M\left( K,\text{A}_{n_{\text{l}}},+2_{n_{\text{l}}} \right) 
\notag \\
&=\left[ e^{iKn_{\text{r}}}+e^{i\phi}e^{iKn_{\text{l}}} \right] M\left( K,\text{A}_0,+2_0 \right) 
\notag \\
&=e^{iKn_{\text{r}}}\left[ 1+e^{i\phi}e^{iKd} \right] M\left( K,\text{A}_0,+2_0 \right) \label{Giant_transition}.
\end{align}
Note that, $\phi$ is the initially encoded phase, and $\Phi = Kd$ is the accumulated propagation phase. With the assistance of a CLS at $n_{\text{l}}$ ($n_{\text{r}}$), the emitter can excite a doublon at $n_{\text{l}}$ ($n_{\text{r}}$) with the center-of-mass $x_c = (n_{\text{r}}+n_{\text{r}})/2$ [$x_c = (n_{\text{l}}+n_{\text{l}})/2$]. Moreover, the propagation phase is associated with $x_c$, and the phase difference between two doublons at $x_{\text{l}/\text{r}}$ is given by $\Phi=Kd$\cite{WangXin2024_2}. For a giant emitter in the single-photon regime, the decay channels depend solely on the emitter coupling legs. In our proposal, the quasi-giant emitter is hybrid of an excited emitter and a CLS. The decay channel is formed by one coupling leg and a CLS. If the CLS associated with one leg is not excited, this leg becomes decoupled, prohibiting emission, and the corresponding emission channel vanishes. For example, if the state $|+2_{n_{\text{r}}}\rangle$ is not excited, the term $M\left( K,\text{A}_{n_{\text{r}}},+2_{n_{\text{r}}} \right)$ in Eq.~\ref{Giant_transition} vanishes. Therefore, we can select the radiation channels by selectively exciting CLS.

\begin{figure*}[t!]
	\centering \includegraphics[width=18cm]{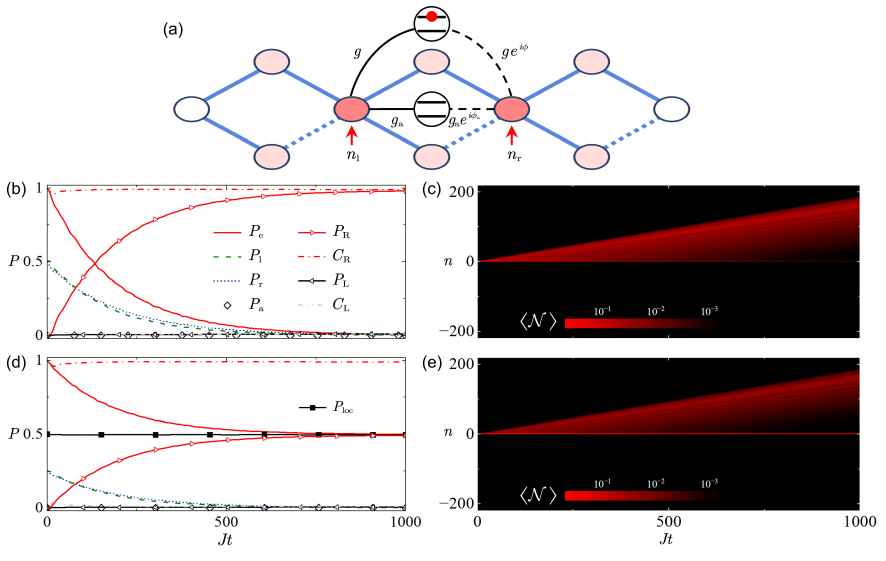}
	\caption{ \textbf{Optimal unidirectional emission}. \textbf{a} Schematic of a giant emitter and an auxiliary emitter coupled to a nonlinear waveguide. Coupling strengths: giant emitter ($g,ge^{i\phi}$) with $g=0.02J$, $\phi =-\pi/2$; auxiliary emitter ($g_{\text{a}},g_{\text{a}}e^{i\phi_{\text{a}}}$) with $g_{\text{a}}=0.072J$, $\phi_{\text{a}}=-\pi/2$. \textbf{b,} \textbf{d} Time evolution of the populations: giant emitter ($P_{\text{e}}$, red solid curves), auxiliary emitter ($P_{\text{a}}$, gray hollow diamonds), compact localized eigenstates ($P_{\text{l}}$, green dashed curves; $P_{\text{r}}$, black dotted curves), left-moving photon field ($P_{\mathrm{L}}$, black curves with left-pointing triangle), right-moving photon field ($P_{\mathrm{R}}$, red curves with right-pointing triangle), two chiral factors ($C_{\mathrm{L}}$, gray dash–double-dot curves; $C_{\mathrm{R}}$, red dash-dotted curves), and localized state ($P_{\mathrm{loc}}= |\psi_{\mathrm{Loc}}|^2$, black curve with squares). \textbf{c,} \textbf{e} Photon number $\langle \mathcal{N} \rangle $ as a function of position $n$ and time $Jt$. Panels (\textbf{b,} \textbf{c}) correspond to the initial state in Eq.~(\ref{psi_full}); panels (\textbf{d,} \textbf{e}) to state in Eq.~(\ref{psi_half}). Other parameters are the same as those in Fig.~\ref{fig4}.} 
	\label{fig5}
\end{figure*}

Similar to the previous derivation, the decay rate is obtained as
\begin{gather}\label{Giant_decay}
\Gamma_{\pm K_r} = \frac{\Gamma _0}{2}|1+e^{i\phi}e^{\pm iK_rd}|^2,
\end{gather}
where $\Gamma_0$ is the decay rate of small quasi emitter in Eq.~(\ref{decay_rate}), and only the relative phase $K_rd$ affects the dynamics. The evolution curves and the photon field distribution are shown in Fig.~\ref{fig4}b,c. The parameters are $d=1$, $\phi=-\pi/2$ and $|K_r|=\pi/2$, which yield $\Gamma_{- K_r} =0 < \Gamma_{+K_r}=\Gamma _0$. Under these conditions, the quasi-giant emitter is expected to exhibit unidirectionally radiation. However, the decay rate of giant emitter matches well, while the radiation field does not display optimal chirality. Below, we explain the origin of this behavior and demonstrate how to realize perfect unidirectional emission by introducing an auxiliary emitter.

In the single-photon regime, since the photon resides in the emitter, realizing perfect chiral emission only requires the effective coupling strength of the giant emitter to exhibit asymmetry and vanish at one directional resonance mode. However, in the nonlinear bath, the situation becomes more complex, necessitating the consideration of the photon state of the bath. As shown in Fig.~\ref{fig4}c, the photon field does not exhibit a perfectly unidirectional distribution. This phenomenon can be attributed to the immobility of CLSs. The system can be modeled as two states $\Psi_{\text{l}}=|e,+2_{n_{\text{l}}}\rangle$ and $\Psi_{\text{r}}=|e,+2_{n_{\text{r}}}\rangle$, coupled to the waveguide, analogous to two entangled emitters. Both states can excite the doublon state, thereby generating an effective interaction mediated by the doublon, which arises from the mutual influence of the radiation field\cite{Sheremet2023}. The effective interaction strength can be derived~\cite{Goban2015,Caneva2015,GonzlezTudela2015,Albrecht2019,Zhang2019}
\begin{gather}
J_{\mathrm{eff}} = g^2M^2\left( K_r,\text{A}_0,+2_0 \right) e^{i\phi}/v_g,
\end{gather}
where $e^{i\phi}$ arises from the encoded phase of the giant emitter. This transition disrupts the unidirectional emission, resulting in a chirality ratio $C_{\mathrm{R}}:C_{\mathrm{L}} \simeq  3:1$, where $C_{\mathrm{R}/\mathrm{L}}=P_{\mathrm{R}/\mathrm{L}}/(P_{\mathrm{R}}+P_{\mathrm{L}})$, and $P_{\mathrm{R}/\mathrm{L}}=\sum_{m,n\gtrless 0}{|\psi _{\mathrm{II}}\left( m,n \right) |^2}$ (see Fig.~\ref{fig4}b). 

Now, we have identified the underlying reason for the suboptimal chirality. To address this disadvantage, we introduce an auxiliary emitter to realize the optimal chiral emission. Note that, the CLEs do not interact with each other, and can be regarded as isolated emitters. By leveraging the concept of dipole-dipole interaction, we introduce a detuned emitter being ground state coupled to two CLSs to facilitate the transition between these CLSs. Furthermore, this setup can establish an auxiliary effective coupling between two states $\Psi_{l/r}$ to counteract the transition mediated by doublon\cite{Guimond2020}. The frequency of the auxiliary emitter is set to $\omega_{\text{e}}^{\text{a}} = 3 $, detuned from the CLS frequency, such that its population can be adiabatically eliminated. The interactions between the auxiliary emitter and two CLSs are $g_{\text{a}}$ and $g_{\text{a}}e^{i\phi_{\text{a}}}$, which counteract the encoded phase of the giant emitter. Consequently, the coupling strength between two CLEs, and thus the states $\Psi_{l/r}$, mediated by the auxiliary emitter, is as follows\cite{Douglas2015,Evans2018}
\begin{gather}
J'_{\mathrm{eff}}=\frac{2g_{\text{a}}^{2}e^{i\phi}}{\omega_{\text{e}}^{\text{a}}-\omega _{\mathrm{CLS}}},
\end{gather}
which is easily obtained via the framework of dipole-dipole interaction, $g^2/\Delta$. Here, $2=|\langle a_{\text{A}_0}\beta _{+2_0}^{\dagger}\rangle |^2$ represents the overlap between the coupling position of the auxiliary emitter and the mode $|+2_0\rangle$. By setting $J_{\mathrm{eff}}=J'_{\mathrm{eff}}$, we plot in Fig.~\ref{fig5}b the time evolution of the populations: quasi-giant emitter ($P_{\text{e}}$), two CLSs $|+2_{n_{\text{l}/\text{r}}}\rangle$ ($P_{\text{l}/\text{r}}$), auxiliary emitter ($P_{\text{a}}$), right and left photon fields ($P_{\mathrm{R/L}}$), and chiral factor ($C_{\mathrm{R/L}}$). In addition, Fig.~\ref{fig5}c,e show the spatial and temporal distribution of the photon number $$\langle \mathcal{N} \left( n,t \right) \,\,\rangle =\langle \Psi \left( t \right) |\sum_{\tau =\text{A,B,C}}{a_{\tau _n}^{\dagger}a_{\tau _n}}|\Psi \left( t \right) \rangle.$$ The population of the auxiliary emitter remains consistently zero, i.e., $P_{\text{a}}(t)=0$, validating the adiabatic elimination approximation. Most importantly, the right chiral factor can reach $C_{\mathrm{R}}=99\%$, enabling optimal unidirectional emission.

Based on triggered emission, the environmental photon state can shape the emitter dynamics. This provides an additional degree of freedom for coherent quantum states manipulation. For example, we demonstrate partial unidirectional emission by exciting two points $\text{A}_{n_{\text{l}}}$ and $\text{A}_{n_{\text{r}}}$. The initial state is 
\begin{align}
&|\psi (t=0)\rangle =|e,g\rangle\! \otimes\! \frac{|\text{A}_{n_{\text{l}}}\rangle +|\text{A}_{n_{\text{r}}}\rangle}{\sqrt{2}}
\notag \\
&=|e,g\rangle\! \otimes\! \left[ \frac{|+\!2_{n_{\text{l}}}\rangle +|+\!2_{n_{\text{r}}}\rangle}{2}\!-\!\frac{|-\!2_{n_{\text{l}}}\rangle +|-\!2_{n_{\text{r}}}\rangle}{2} \right]  \label{psi_half}
\end{align}
As shown in Fig.~\ref{fig5}d,e, the time evolution of the populations and photonic field distribution reveals that the system evolves into a coherent superposition:
\begin{gather}
|\psi (t_{\text{f}})\rangle =\frac{1}{\sqrt{2}}\left[ |g,g,\psi _{\exp}^{\text{r}}\rangle +|e,g,\psi _{\mathrm{loc}}\rangle \right].
\end{gather}
Here, $\psi _{\mathrm{loc}}\!=\!\left(|-\!2_{n_{\text{l}}}\rangle \!+\!|-\!2_{n_{\text{r}}}\rangle\right)/\sqrt{2} $ denotes the localized component that remains trapped due to energy mismatch; while $\psi _{\exp}^{\text{r}}$ is a unidirectional right-going doublon wave packet. Therefore, by engineering the initial single-photon state, one can selectively control the fraction of the state which exhibits unidirectional radiation.

Two giant emitters can also chirally excite the doublon state and unidirectionally radiate, without involving any additional operators~\cite{WangXin2024_2}. This is because the single-photon state arises from an intermediate state which is virtually excited by the emitter. The single-photon states at two coupling positions are inherently connected, analogous to the double-slit interference, where a single source corresponds to two spatial positions. 

Furthermore, based on this chiral emission, the two correlated photons can act as flying qubits, enabling the transfer of more information between remote nodes\cite{Lodahl2017,WangXin2024_2}. By modulating the setup, the unidirectionally emitted doublon  from a quasi-giant emitter can be reabsorbed by another quasi-giant emitter, where one photon excites the emitter and the other excites the corresponding CLS state. Alternatively, this proposal can also achieve the propagation of localized states within a flat-band system. These applications demonstrate that, in the nonlinear bath, although light-matter interactions retain some similarities with the linear case, they are significantly more complex and exhibit distinct characteristics. The quantum state of the environment becomes increasingly critical, necessitating careful consideration.

\vspace{.3cm}
\noindent {\textbf{Implementation}}
\\
\noindent
Our proposal is experimentally feasible using superconducting circuit-QED platforms. The flexible connectivity between transmon qubits allows for the construction of various lattice geometries, while parametric modulation of coupling strengths enables the synthesis of effective gauge fields with tunable hopping phases~\cite{Roushan2016,Rosen2024,Rosen2024_2}. The required nonlinear local potential is naturally provided by the intrinsic anharmonicity of transmons~\cite{Koch2007,Wang2024}. A recent experiment has realized such all-bands-flat system with strong photon-photon interactions in superconducting circuits, where interaction-induced delocalization was observed~\cite{Martinez2023}. Typically~\cite{Rosen2024_2,Wang2024,Mansikkamaki2022}, the lattice frequency is set $\omega_c/2\pi \sim 5$ GHz, with an anharmonicity $U/2\pi \sim 160$ MHz, and the hopping rate $J/2\pi \sim 40$ MHz, consistent with the condition $U = 4J$ used in our theoretical analysis.

Furthermore, giant emitters with multiple coupling points~\cite{Kannan2020} and engineered coupling phases via parametric modulation~\cite{Joshi2023} have already been demonstrated in circuit-QED architectures. In our setup, the emitter qubit has an anharmonicity of $U_{\text{e}}/2\pi \sim 300$ MHz to enable two-level behavior and is capacitively coupled to a lattice site~\cite{Pakkiam2023}. The triggered emission rate is estimated as $\Gamma/2\pi \sim 1$ MHz, which significantly exceeds the typical intrinsic decay ($\gamma/2\pi \sim 10$ kHz) and dephasing rates ($\kappa/2\pi \sim 5$ kHz), satisfying $\Gamma \gg \{\gamma, \kappa\}$~\cite{Kjaergaard2020}. Therefore, the predicted phenomena are resolvable within the coherence times of current devices, enabling near-term experimental verification.

\vspace{.3cm}
\noindent {\large\textbf{Conclusions}}
\\
\noindent
In summary, we demonstrate that in a nonlinear photonic bath, the radiation dynamics of emitters are profoundly influenced by the quantum state of environment. A key discovery is triggered emission, where a far-detuned emitter (unable to radiate independently) is triggered by the environmental photon state to excite highly correlated photon pair doublon. To generalize this framework, we conceptualize the emitter and its associated CLS as a combined quasi emitter, offering a versatile platform for quantum state engineering via tailored initial environmental photonic states. This approach enables the generation of entangled superpositions and partial unidirectional photon transport, with potential applications in quantum information processing.
	
Our findings highlight that in nonlinear bath, the quantum state of the environment can actively shape emitter dynamics, enabling environment-programmable quantum phenomena. This framework, which places the environmental photonic bath at the center of the dynamics, opens a pathway for exploring nonlinear quantum optics. Furthermore, beyond doublons, higher-order correlated states (e.g., triplons)~\cite{Mansikkamaki2022} offer rich opportunities for future exploration in the largely uncharted regime of nonlinear multi-photon dynamics.

Although our present study utilizes a specific gauge-field-induced flat-band system, it raises a fundamental question: does the triggered emission mechanism persist in more realistic disordered systems with Anderson localization~\cite{Lahini2008,Lagendijk2009,Segev2013}? Moreover, our findings motivate investigating analogous phenomena in conventional nonlinear waveguides~\cite{Winkler2006} and other systems with photon-photon interactions~\cite{Chang2014}. We hope this work inspires future studies on triggered emission in a wider variety of nonlinear platforms.

\vspace{.3cm}
\noindent {\large \textbf{Methods}}
\\
\noindent \textbf{Numerical simulation}
\\
\noindent
We simulate the physical system in real space. The Hilbert space is restricted to the two-excitation subspace, i.e., 
\begin{gather}
|\Psi \left( t \right) \rangle\! \!=\!\!\sum_{\tau ,n}\!{c_{e,\tau _n}\!\left( t \right)\! |e,\tau _n\rangle}\!+\!\!\!\sum_{\tau ,n,\tau ',n'}\!\!{c_{\tau _n,\tau '_{n'}}\left( t \right) |\tau _n,\tau '_{n'}\rangle}.
\end{gather}
Here $c_{e,\tau _n}$ denotes the excitation of the emitter and the lattice site $\tau_n$, while $c_{\tau _n,\tau ^{\prime}_{n^{\prime}}} $ represents the excitation of two lattice site $\tau_n$ and $\tau ^{\prime}_{n^{\prime}}$. The full-space Hamiltonian can be expanded in this basis. By numerically solving the Schr\'odinger equation under the initial state, we obtain the probabilities of the emitter $|c_{e,\tau _n}|^2$ and the two-photon field $|c_{\tau _n,\tau ^{\prime}_{n^{\prime}}}|^2$. 

The numerical simulations of quantum dynamics are based on the open-source Python package QuTiP~\cite{Johansson12qutip,Johansson13qutip} and QuSpin \cite{Weinberg2017,Weinberg2019}, which agree well with theoretical results throughout our work.

\vspace{.3cm}
\noindent \textbf{Effective transition rate}
\\
\noindent
The transition rate between two two-photon states can be expressed as 
\begin{align} 
&M\left( K,\text{A}_n,+2_m \right) =\langle \beta _{+2_m}a_{\text{A}_n}\mathcal{D} _{1,K}^{\dagger}\rangle 
\notag \\
&=\! \langle \left(\! 2a_{\text{A}_m} \!\! + \! a_{\text{B}_m} \! - \! a_{\text{C}_m} \! \! + \! a_{\text{B}_{m+1}} \! \! + \! a_{\text{C}_{m+1}} \! \right) a_{\text{A}_n}\mathcal{D} _{1,K}^{\dagger}\rangle. \label{transition_rate_M} 
\end{align}
This quantity represents the overlap between the two-photon doublon mode $\mathcal{D} ^{\dagger}_{1,K}|\mathrm{vac}\rangle$ and a single-photon CLS state combined with a single-point excitation (the emitter exciting) $\beta^{\dagger}_{+2_{n_0}}a^{\dagger}_{\tau_{n_{0}}}|\mathrm{vac}\rangle$. We take an example as 
\begin{gather}
\langle a_{\tau _m}a_{\text{A}_n}\mathcal{D} _{1,K}^{\dagger}\rangle =e^{iK\frac{m+n}{2}}\psi \left( \tau _m\text{A}_n \right) .
\end{gather}
For simplicity, we set $n=0$. As shown in Fig.~\ref{fig1}e in the main text, owing to the compact property, the doublon only has the distribution at the Fock state $|\text{A}_0\text{A}_0\rangle$, $|\text{A}_0\text{B}_{0/1}\rangle$, $|\text{A}_0\text{C}_{0/1}\rangle$ with $\text{A}_0$ being excited. Therefore, we obtain the condition
\begin{gather} \label{condition}
\langle a_{\tau _m}a_{\text{A}_0}\mathcal{D} _{1,K}^{\dagger}\rangle \ne 0,\quad \tau _m=\text{A}_0,\text{B}_{0/1},\text{C}_{0/1}.
\end{gather}
We further analyze the transition rate Eq.~(\ref{transition_rate_M}) with different $m$

1) For $|m|>1$, the eigenmode $\beta _{+2_m}$ does not have distribution in position $\tau _m=\text{A}_0,B_{0/1},C_{0/1}$, leading to the overlap vanishing, $M=0$. 

2) For $m=1$, Eq.~(\ref{transition_rate_M}) is 
\begin{align}
&M\left( K,\text{A}_0,+2_1 \right) =\langle \beta _{+2_1}a_{\text{A}_0}\mathcal{D} _{1,K}^{\dagger}\rangle 
\notag \\
&=\langle \left( 2a_{\text{A}_1}+a_{\text{B}_1}-a_{\text{C}_1}+a_{\text{B}_2}+a_{\text{C}_2} \right) a_{\text{A}_0}\mathcal{D} _{1,K}^{\dagger}\rangle 
\notag \\
&=\langle \left( a_{\text{B}_1}-a_{\text{C}_1} \right) a_{\text{A}_0}\mathcal{D} _{1,K}^{\dagger}\rangle =0.
\end{align} 
Here, only $a_{\text{B}_1}$ and $a_{\text{C}_1}$ satisfy the condition in Eq.~(\ref{condition}). However, the single-photon states $+a_{\text{B}_1}$ and $-a_{\text{C}_1}$ have identical value, but opposite phases, while the doublon states $\psi \left( \text{A}_0\text{B}_0 \right) = \psi \left( \text{A}_0\text{C}_0 \right)$ have identical values and phases. This results in the transition rate being zero, as discussed in the main text. 

3) For $m=-1$, Eq.~(\ref{transition_rate_M}) becomes
\begin{align}
&M\left( K,\text{A}_0,+2_{-1} \right) =\langle \beta _{+2_{-1}}a_{\text{A}_0}\mathcal{D} _{1,K}^{\dagger}\rangle \notag \\
&=\langle \left( 2a_{\text{A}_{-1}}+a_{\text{B}_{-1}}-a_{\text{C}_{-1}}+a_{\text{B}_0}+a_{\text{C}_0} \right) a_{\text{A}_0}\mathcal{D} _{1,K}^{\dagger}\rangle \notag \\
&=\langle \left( a_{\text{B}_0}+a_{\text{C}_0} \right) a_{\text{A}_0}\mathcal{D} _{1,K}^{\dagger}\rangle =0.
\end{align} 
In this case, the single-photon states $+a_{\text{B}_0}$ and $+a_{\text{C}_0}$ have identical values and phases, while the doublon state $\psi \left( \text{A}_0\text{B}_0 \right) = - \psi \left( \text{A}_0\text{C}_0 \right)$ have identical values, but opposite phases. This leads to the transition rate remaining zero. 

Ultimately, the transition rate Eq.~(\ref{transition_rate_M}) is non-zero only when $m=n$, and only the term $M\left( K, \tau_{n_{0}} ,+2_{n_{0}} \right)$ survives, while all others terms $M\left( K, \tau_{m} ,+2_{n_{0}} \right)$, $m\ne n_0$ vanish. Thus, we have
\begin{gather}
\sum_{m}M\left( K,\text{A}_{n_0},+2_m \right) = M\left( K,\text{A}_{n_0},+2_{n_0} \right).
\end{gather}
Although the gauge field induces the asymmetry of the single- and doublon-wavefunction, the transition coefficients precisely combine both contributions and cancel out the asymmetries, yielding single point distributed transition coefficients.

\vspace{.3cm}
\noindent {\large \textbf{Data availability}}
\\
\noindent
Numerical source data for all the figures are provided at ~\url{https://zenodo.org/records/17404329}. Additional data that support the findings of this study are available from the authors upon reasonable request.


\vspace{.3cm}
\noindent {\large \textbf{Code availability}}
\\
\noindent
The codes used for the simulation and analysis are available from the authors upon reasonable request.

\vspace{.3cm}
\noindent {\large \textbf{References}}
\\
\noindent

\bibliography{Triggered_emission_ref}

\vspace{.3cm}
\noindent{\large \textbf{Acknowledgements} 
\\
\noindent
X.W.~is supported by the National Natural Science Foundation of China (NSFC) (Grant No.~12174303).

\vspace{.3cm}
\noindent{\large \textbf{Author contributions} 
\\
\noindent
J.Q.L. and X.W. conceived the original idea. J.Q.L. did the analytical and numerical analysis under the supervision of X.W. J.Q.L. and X.W. wrote the manuscript. 

\vspace{.3cm}
\noindent{\large \textbf{Competing interests} 
\\
\noindent
The authors declare no competing interests.

\vspace{.3cm}
\noindent{\large \textbf{Supplementary information} 
\\
\noindent
The online version contains supplementary material available at

\end{document}